\documentclass{article}%
\usepackage{amsfonts}
\usepackage{amsmath}
\usepackage{amssymb,amsbsy}
\usepackage{graphicx}%
\begin{document}

\title{\bf Bi-Hamiltonian representation of St\"{a}ckel systems}
\author{Maciej B\l aszak \\
Faculty of Physics, A. Mickiewicz University,\\Umultowska 85,
61-614 Pozna\'{n}, Poland\\
} \maketitle

\begin{abstract}
It is shown that a linear separation relations are fundamental
objects for integration by quadratures of St\"{a}ckel separable Liouville integrable systems (the so-called St\"{a}ckel systems).
These relations are further employed for the classification of St\"{a}ckel
systems. Moreover, we prove that {\em any}
St\"{a}ckel separable Liouville integrable system
can be lifted to a bi-Hamiltonian system of Gel'fand-Zakharevich type.
In conjunction with other known result this
implies that the existence of bi-Hamiltonian
representation of Liouville integrable systems is a necessary condition for St\"{a}ckel separability.
\end{abstract}

\section{Introduction}

The Hamilton-Jacobi (HJ) theory seems to be one of the most powerful methods
of integration by quadratures for a wide class of systems described by
nonlinear ordinary differential equations, with a long history as a
part of analytical mechanics. The theory in question is closely related to the
Liouville integrable Hamiltonian systems. The milestones
of this theory include the works of St\"{a}ckel, Levi-Civita, Eisenhart,
Woodhouse, Kalnins, Miller, Benenti and others. The majority of
results was obtained for a very special class of integrable systems,
important from the physical point of view, namely for the systems with
quadratic in momenta first integrals.

The first efficient construction of the separation variables for dynamical systems was
discovered by Sklyanin \cite{sk1}. He adapted the methods of soliton theory, i.e., the
Lax representation and r-matrix theory for systematic derivation of separation
coordinates. In this approach the integrals of motion in involution appear as coefficients of
characteristic equation (\emph{spectral curve}) of the Lax matrix. This method was
successfully applied for separating variables in many integrable systems
\cite{sk1}-\cite{ma1}.

Recently, a modern geometric theory of separability on bi-Poisson manifolds
was developed \cite{1}-\cite{m3}. This theory is closely related to the so-called
Gel'fand-Zakharevich (GZ) bi-Hamiltonian systems \cite{GZ},\cite{GZ1}.
The theory in question includes Liouville integrable systems
with integrals of motion being functions quadratic in momenta as a very special case. In this
approach the constants of motion are closely related to the so-called
\emph{separation curve} which is intimately related to the St\"{a}ckel
separation relations. The separation curve arising
in the geometric approach is closely related to its counterpart in the r-matrix
approach. In fact, these curves are identical for linear r-matrix and related by
exponentiation of momenta in the spectral curve for dynamical (quadratic) r-matrix
\cite{kuz}, \cite{8}.

In the present paper we develop in a systematic fashion a separability theory
of the Liouville integrable systems which are of the GZ type, including as a
special case the class of systems with quadratic in momenta first integrals.
First of all, we treat St\"{a}ckel separable systems according to the
form of separation relations and make some observations related to their
classification. Then we construct a quasi-bi-Hamiltonian representation of
St\"{a}ckel systems\ on $2n$ dimensional phase space and lift them to the
related GZ bi-Hamiltonian systems on the extended $2n+k$ dimensional phase space. This
result proves that bi-Hamiltonian property is common to all classes of the
St\"{a}ckel systems considered. In other words, we prove that the existence of bi-Hamiltonian
representation for Liouville integrable systems is a necessary condition for their St\"{a}ckel separability, i.e.
for these systems for which separation relations are linear in all constants of motion which are in involution.
Finally let us mention that up to now such a proof was available only for a
distinguished class of the so-called Benenti systems \cite{ib}, where $k=1$
and all constants of motion are quadratic in momenta.

%\newpage
\section{\bigskip Separable St\"{a}ckel systems}

Consider a Liouville integrable system on a $2n$-dimensional phase space
$M$. Thus, we have $M\ni u=(q_{1},\dots,q_{n},p_{1},\dots,p_{n})^{T}$ and
there are $n$ functions
$H_{i}(q,p)$ in involution with respect to the canonical Poisson
tensor $\pi$
\[
\{H_{i},H_{j}\}_{\pi}=\pi(dH_{i},dH_{j})=\langle dH_{i},\pi\,dH_{j}\rangle
>=0,\qquad i,j=1,\dots,n,
\]
where $\langle\cdot,\cdot\rangle$ is the standard pairing of $TM$ and $T^{\ast}M$.
Canonicity of $\pi$ means that the only nonzero Poisson brackets among the
coordinates are $\{q_{i},p_{j}\}_{\pi}=\delta_{ij}$. The functions $H_{i}$
generate $n$ Hamiltonian dynamic systems
\begin{equation}
u_{t_{i}}=\pi\,dH_{i}=X_{H_{i}},\qquad i=1,\dots,n, \label{1}%
\end{equation}
where $X_{H_{i}}$ are called the \emph{Hamiltonian vector fields}.

The Hamilton-Jacobi (HJ) method for solving (\ref{1}) essentially amounts to the
linearization of the latter via a canonical transformation
\begin{equation}
(q,p)\rightarrow\ (b,a),\quad a_{i}=H_{i},\qquad i=1,\dots,n. \label{2}%
\end{equation}
In order to find the conjugate coordinates $b_{i}$ it is necessary to
construct a generating function $W(q,a)$ of the transformation (\ref{2}) such
that
\[
b_{j}=\tfrac{\partial W}{\partial a_{j}},\quad p_{j}=\tfrac{\partial W}{\partial
q_{j}}.
\]

The function $W(q,a)$ is a complete integral of the associated
\emph{Hamilton-Jacobi equations}
\begin{equation}
H_{i}\left(  q_{1},\dots,q_{n},\tfrac{\partial W}{\partial
q_{1}},\dots,\tfrac
{\partial W}{\partial q_{n}}\right)  =a_{i},\qquad i=1,\dots,n. \label{3}%
\end{equation}
In the $(b,a)$ representation the $t_{i}$-dynamics is trivial:
\[
(a_{j})_{t_{i}}=0,\quad (b_{j})_{t_{i}}=\delta_{ij},
\]
whence
\begin{equation}
b_{j}(q,a)=\tfrac{\partial W}{\partial a_{j}}=t_{j}+c_{j},\qquad j=1,\dots,n, \label{4}%
\end{equation}
where $c_j$ are arbitrary constants.

Equations (\ref{4}) provide implicit solutions for (\ref{1}). Solving them for $q_j$ is
known as the \emph{inverse Jacobi problem}. The reconstruction in explicit form of
trajectories $q_{j}=q_{j}(t_{i})$ is in itself a highly nontrivial problem from algebraic geometry
which is beyond the scope of this paper.

%Then, one can ask the question where are the difficulties in that method. The
The main difficulty in applying the above method to a given Liouville integrable system in given
canonical coordinates $(q,p)$ consists in solving the system (\ref{3}) for $W$. In general this is a
hopeless task, as (\ref{3}) is a system of nonlinear coupled partial
differential equations. In essence, the only hitherto known way
of overcoming this difficulty is to find
distinguished canonical coordinates, denoted here by $(\lambda,\mu)$, for
which there exist $n$ relations
\begin{equation}
\varphi_{i}(\lambda_{i},\mu_{i};a_{1},\dots,a_{n})=0,\qquad i=1,\dots,n,\ a_{i}%
\in\mathbb{R}\mathbf{,}\quad \det\left[  \tfrac{\partial\varphi_{i}}{\partial
a_{j}}\right]  \neq 0, \label{5}%
\end{equation}
such that each of these relations involves only a single
pair of canonical coordinates \cite{sk1}. The determinant
condition in (\ref{5}) means that we can solve the equations (\ref{5}) for $a_{i}$
and express $a_i$ in the form $a_{i}=H_{i}(\lambda,\mu)$, $i=1,\ldots,n$.

If the functions $W_{i}(\lambda_{i},a)~$\ are solutions of a system of $n$
{\em decoupled} ODEs obtained from (\ref{5}) by substituting $\mu_{i}%
=\tfrac{dW_{i}(\lambda_{i},a)}{d\lambda_{i}}$%
\begin{equation}
\varphi_{i}\left(  \lambda_{i},\mu_{i}=\tfrac{dW_{i}(\lambda_{i}%
,a)}{d\lambda_{i}},a_{1},\ldots,a_{n}\right)  =0,\quad i=1,\dots,n,
\label{SRdiff}%
\end{equation}
then the function
\[
W(\lambda,a)=\sum\nolimits_{i=1}^{n}
W_{i}(\lambda_{i},a)
\]
is an additively separable solution of \emph{all} the
equations (\ref{SRdiff}), and \emph{simultaneously} it is a solution of all
Hamilton-Jacobi equations (\ref{3}) because solving (\ref{5}) to the
form $a_{i}=H_{i}(\lambda,\mu)$ is a purely algebraic operation. The
Hamiltonian functions $H_{i}$ Poisson commute since the constructed function
$W(\lambda,a)$ is a generating function for the canonical transformation
$(\lambda,\mu)\rightarrow(b,a)$. The distinguished coordinates $(\lambda,\mu)$ for which the original
Hamilton-Jacobi equations (\ref{3}) are equivalent to a set of separation relations
(\ref{SRdiff}) are called the
\emph{separation coordinates}.

Of course, the original Jacobi formulation of the method was a bit different
from the one presented above, and was made for a particular class of
Hamiltonians, nevertheless it contained all important ideas of the method.
Jacobi himself doubted whether there exists a systematic method for
construction of separation coordinates. Indeed, for many decades of development
of separability theory, the method did not exist. Only recently, at the end of
the $20$th century, after more then one hundred years of efforts, two
different constructive methods were suggested, the first related to the Lax
representation and the second related to the bi-Hamiltonian representation for a given
integrable system.

We would like to stress that all results of the present paper
are derived directly from the separation relations
(\ref{5}), thus confirming their fundamental role in the modern separability theory.

In what follows we restrict ourselves to considering a special case of (\ref{5}) when all
separation relations are affine in $H_{i}$:
\begin{equation}
\sum_{k=1}^{n}S_{i}^{k}(\lambda_{i},\mu_{i})H_{k}=\psi_{i}(\lambda_{i},\mu
_{i}),\qquad i=1,\dots,n, \label{6}%
\end{equation}
where $S^k_i$ and $\psi_i$ are arbitrary smooth functions of their arguments.
The relations (\ref{6}) are called
the generalized \emph{St\"{a}ckel separation relations} and the related
dynamical systems are called the \emph{St\"{a}ckel separable} ones. The matrix $S=(S^k_i)$
will be called a \emph{generalized St\"{a}ckel matrix}. The reason
behind this name is the fact that the conditions (\ref{6}) with $S_{i}^{k}$ being $\mu
$-independent and $\psi_{i}$ being quadratic in momenta $\mu$ are equivalent
to the original St\"{a}ckel conditions for separability of Hamiltonians
$H_{i}$. To recover the explicit St\"{a}ckel form of the Hamiltonians it suffices
to solve the linear system (\ref{6}) with respect to $H_{i}$.

Although the restriction of linearity appears to be very strong,
for all known separable systems (at least to the knowledge of the author), the general separation
conditions can be reduced to the form (\ref{6}) upon suitable choice of integrals
of motion $H_{i}.$ The possible explanation of this fact is that we simply have no mathematical tools
for effective construction of separation coordinates for non-St\"{a}ckel separable systems, so that part of
separability theory is yet terra incognita.

Let us come back to the St\"{a}ckel case. If in (\ref{6}) we further have $S_{i}^{k}(\lambda_{i},\mu_{i}%
)=S^{k}(\lambda_{i},\mu_{i})$ and $\psi_{i}(\lambda_{i},\mu_{i})=\psi
(\lambda_{i},\mu_{i})$ then the separation conditions can be represented by $n$
copies of the curve
\begin{equation}
\sum_{k=1}^{n}S^{k}(\lambda,\mu)H_{k}=\psi(\lambda,\mu) \label{7}%
\end{equation}
in $(\lambda,\mu)$ plane, called a \emph{separation curve}. The copies in question are obtained by setting
$\lambda=\lambda_i$ and $\mu=\mu_i$ for $i=1,\dots,n$.

\textbf{Remark.} There is an important special case when (\ref{7})
is an arbitrary nonsingular compact Riemann surface $\Gamma$, i.e., when
$S^{k}(\lambda,\mu)$ and $\psi(\lambda,\mu)$ are polynomials of
$\lambda$ and $\mu$ of certain specific form. Then one can find the
genus of this curve and basic holomorphic differentials in a
standard fashion and the Jacobi inversion problem (\ref{4}) can be
equivalently expressed by the Abel map of the Riemann surface
$\Gamma$ into its Jacobi variety and solved in the language of
Riemann theta functions (see \cite{dkn} and references therein).

From now on we
will consider St\"{a}ckel separable systems with separation relations of
the most general form (\ref{6}). For reasons to be explained in the next section, we collect
the terms from the l.h.s. of (\ref{6}) as follows:
\begin{equation}
\sum_{k=1}^{m}\varphi_i^k(\lambda_i,\mu_i)H^{(k)}(\lambda_i)=\psi_i
(\lambda_i,\mu_i),\qquad i=1,\dots,n,
\label{9}%
\end{equation}
where
\[
H^{(k)}(\lambda)=\sum_{i=1}^{n_{k}}\lambda^{n_{k}-i}H_{i}^{(k)}%
,\qquad n_{1}+\dots+n_{m}=n
\]
and impose the normalization $\varphi_i^m(\lambda_i,\mu_i)=1$.

As the separation relations (\ref{6}) play the fundamental role in the
Hamilton-Jacobi theory, it is natural to employ them for
classification of St\"{a}ckel systems. The form of separation
relations (\ref{9}) allows us to classify the associated St\"{a}ckel
systems. Actually, any given class of St\"{a}ckel separable systems
can be represented by a fixed St\"{a}ckel matrix $S$ and the functions $\psi$.
The matrix $S$ is uniquely defined by $m$ vectors
$\varphi^k=(\varphi^k_1,\dots,\varphi^k_n)^T$, $k=1,\dots,m$, and the
partition $(n_1,\dots,n_m)$ of $n$. Note that in our normalization we have
$\varphi^m=(1,\dots,1)^T$.

For example, the most
intensively studied systems in the 20th century, those related to
one-particle separable dynamics on Riemannian manifolds with flat or constant
curvature metrics, belong to the simplest class with
$m=1$ and the functions $\psi_i$ being quadratic in the momenta $\mu_i$:
\begin{equation}
\sum_{j=1}^nH_j\lambda_i^{n-j}=\tfrac{1}{2}f_i(\lambda_i)\mu^{2}_i
+\gamma_i(\lambda_i),\qquad i=1,\dots,n. \label{ben}
\end{equation}
This class, which will hereinafter be
referred to as the Benenti class, includes systems generated by conformal Killing
tensors \cite{be}-\cite{ib}, as well as bi-cofactor systems, generated by a
pair of conformal Killing tensors \cite{cof1}-\cite{mk}. Here the functions $f_i$ define
the St\"{a}ckel metric while the functions $\gamma_i$ define a separable potential.
When $f_i=f(\lambda_i)$ and $f$ is a polynomial of order not higher then $n+1$, then the associated
St\"{a}ckel metric is of constant curvature.

Another class of separable systems also has $m=1$
but the functions $\psi_i$ are now exponential in
the momenta
\[
\sum_{j=1}^nH_j\lambda_i^{n-j}=\exp(a\mu_i)+\exp(-b\mu_i
)+\gamma_i(\lambda_i),\qquad i=1,\dots,n,
\]
where $\gamma_i$ define a separable potential.
This class includes such systems as the periodic Toda lattice
\cite{m2}, the KdV dressing chain \cite{8}, the Ruijsenaars-Schneider system
\cite{7} and others.

We also know some particular examples from the
classes with $m>1$. For instance, stationary flows of the Boussinesq hierarchy belong to
the class with $m=2$, $n_1=1$, $n_2=n-1$, $\varphi^1_i=\mu_i$ \cite{m1}, \cite{7}. Dynamical system
on loop algebra $\widehat{\mathfrak{sl}}(3)$ belongs to the class with $m=2$,
$n_1=2$, $n_2=4$, $\varphi^1_i=\mu_i$ \cite{m3}. In both cases the functions
$\psi_i$ are cubic in the momenta, so these separation relations belong to the following class:
\begin{equation}
\mu_i\sum_{j=1}^{n_1} H_j^{(1)} \lambda_i^{n_1-j}+\sum_{j=1}^{n_2} H_j^{(2)}
\lambda_i^{n_1-j}=\tfrac{1}{3}f(\lambda_i)\mu_i^{3}
+\mu_i \gamma_1(\lambda_i)+\gamma_2(\lambda_i),\qquad i=1,\dots,n, \label{bus}
\end{equation}
where $\mu \gamma_1$ and $ \gamma_2$ give rise to the separable potentials.

Finally, systems from the classes with $1<m\leq n$, $\varphi^k_i=\lambda_i^{(\alpha_{k})}$,
$\alpha_k\in\mathbb{N}$ and with $\psi_i$ quadratic in the momenta, i.e.
\begin{equation}
\sum_{k=1}^{m}\lambda_i^{(\alpha_{k})}H^{(k)}(\lambda_i)=\tfrac{1}{2}f_i(\lambda_i)\mu^{2}_i
+\gamma_i(\lambda_i),\qquad i=1,\dots,n,
\label{99}%
\end{equation}
 were
constructed in \cite{mac2005} and \cite{bs}.

\section{Bi-Hamiltonian property of St\"{a}ckel systems}

We start this section with a few definitions important for further
considerations. As the Hamiltonian formalism is of tensorial type,
there is no need to restrict ourselves to nondegenerate
canonical representation of Hamiltonian vector fields. Given a
manifold $\mathcal{M}$, a \emph{Poisson operator} $\pi$ on
$\mathcal{M}$ is a bivector (second order contravariant tensor
field) with vanishing Schouten bracket
\begin{equation*}
\lbrack\pi,\pi\rbrack_{S}=0. \label{0.2}%
\end{equation*}
Then the bracket
\[
\{f_{1},f_{2}\}_{\pi}:=\langle df_{1},\pi df_{2}\rangle,\qquad f_{1},f_{2}\in C^\infty(\mathcal{M}),
%:\mathcal{M}\rightarrow\mathbb{R},
\]
is the Lie bracket, i.e., it is skew-symmetric and satisfies the Jacobi identity.
A function $c:\mathcal{M}\rightarrow\mathbb{R}$ is
called the \emph{Casimir function} of the Poisson operator $\pi$ if for an
arbitrary function $f:\mathcal{M}\rightarrow\mathbb{R}$ we have $\left\{
f,c\right\}  _{\pi}=0$ (or, equivalently, if $\pi dc=0$). A linear combination
$\pi_{\lambda}=\pi_{1}-\lambda\pi_{0}$ ($\lambda\in\mathbb{R}$) of two Poisson
operators $\pi_{0}$ and $\pi_{1}$ is called a \emph{Poisson pencil} if the
operator $\pi_{\lambda}$ is Poisson for any value of the parameter $\lambda$,
i.e., when $[\pi_{0},\pi_{1}]_{S}=0$. In this case we say that $\pi_{0}$ and
$\pi_{1}$ are \emph{compatible}. Given a Poisson pencil $\pi_{\lambda}%
=\pi_{1}-\lambda\pi_{0}$ we can often construct a sequence of vector fields
$X_{i}$ on $\mathcal{M}$ that have two Hamiltonian representations (the so-called
\emph{bi-Hamiltonian chain})
\begin{equation}
X_{i}=\pi_{1}dh_{i}=\pi_{0}dh_{i+1}, \label{0.3}%
\end{equation}
where %$h_{i}:\mathcal{M}\rightarrow\mathbb{R}$
$h_{i}\in C^\infty(\mathcal{M})$ are called the Hamiltonians of
the chain (\ref{0.3}) and where $i\ $\ is a discrete index. This sequence
of vector fields may or may not terminate in zero depending on the existence of the
Casimir functions for the pencil.

Consider a bi-Poisson manifold $(M,\pi_{0},\pi_{1})$ of $\dim M=2n+m$,
where $\pi_{0},\pi_{1}$ is a pair of compatible Poisson tensors of rank $2n$.
We further assume that the Poisson pencil $\pi_{\lambda}$ admits $m$
Casimir functions which are polynomial in the pencil parameter $\lambda$
and have the form
\begin{equation}
h^{(j)}(\lambda)=\sum_{i=0}^{n_{j}}\lambda^{n_{j}-i}h_{i}^{(j)}%
,\qquad j=1,\dots,m, \label{0.4}%
\end{equation}
so that $n_{1}+\dots+n_{m}=n$ and $h_{i}^{(j)}$ are functionally
independent. The collection of $n$ bi-Hamiltonian vector fields
\begin{equation}
\pi_{\lambda}dh^{(j)}(\lambda)=0\Longleftrightarrow X_{i}^{(j)}=\pi_{1}%
dh_{i}^{(j)}=\pi_{0}dh_{i+1}^{(j)},\quad i=1,\dots,n_{j},\quad
j=1,\dots,m,
\label{0.5}%
\end{equation}
is called the Gel'fand-Zakharevich (GZ) system of the bi-Poisson manifold
$\mathcal{M}$. Notice that each chain starts from a Casimir of $\pi_{0}$ and
terminates with a Casimir of $\pi_{1}$. Moreover, all $h_{i}^{(j)}$ pairwise
commute with respect to both Poisson structures
\begin{equation*}
X_{i}^{(j)}(h_{l}^{(k)})=\langle dh_{l}^{(k)},\pi_{0}dh_{i+1}^{(j)}%
\rangle=\langle dh_{l}^{(k)},\pi_{1}dh_{i}^{(j)}\rangle=\{h_{l}^{(k)},h_{i+1}^{(j)}\}_{\pi_0}
=\{h_{l}^{(k)},h_{i}^{(j)}\}_{\pi_1}=0. \label{0.6a}%
\end{equation*}

In the following section we prove that an arbitrary St\"{a}ckel system on the
phase space $M$ with separation conditions given by (\ref{9}) can be lifted
to a GZ bi-Hamiltonian system on the extended phase space $\mathcal{M}$.
\looseness=-1

As recently proved in \cite{m3}, the St\"{a}ckel Hamiltonians from separation relations (\ref{6}) admit
the following quasi-bi-Hamiltonian representation
\begin{equation}
\Pi_{1}dH_{i}=\sum_{j=1}^{n}F_{ij}\,\Pi_{0}\,dH_{j},\qquad
i=1,\dots,n,
\label{0.7}%
\end{equation}
where $\Pi_{0}$ is a canonical Poisson tensor
\[
\Pi_{0}=\left(
\begin{array}
[c]{cc}%
0 & I_{n}\\
-I_{n} & 0
\end{array}
\right),
\]
$I_n$ is an $n\times n$ unit matrix,
$\Pi_{1}$ is a noncanonical Poisson tensor of the form
\[
\Pi_{1}=\left(
\begin{array}
[c]{cc}%
0 & \Lambda_{n}\\
-\Lambda_{n} & 0
\end{array}
\right),\quad
\Lambda_{n}=\mathrm{diag}(\lambda_{1},\dots,\lambda_{n}),
\]
compatible with $\Pi_{0}$, and the \emph{control matrix} $F$ has the form
\begin{equation}
F=(S^{-1}\Lambda_{n}S), \label{0.8}%
\end{equation}
where $S$ is the associated St\"{a}ckel matrix.

To have a better insight into the functions $F_{ij}$,
we will find another
representation for the entries $F_{ij}=(S^{-1}\Lambda_{n}S)_{ij}$ of $F$.
To this end
consider a system of $n$ linear
equations for $V_{k},k=1,\dots,n$;
\begin{equation}
\sum_{k=1}^{n}S^{k}_i(\lambda_i,\mu_i)V_{k}=\sum_{j=1}^{n}\lambda_i
S^{j}_i(\lambda_i,\mu_i)a_{j},\qquad i=1,\dots,n, \label{0.9}%
\end{equation}
where $a_i,i=1,\dots,n$ are some parameters. The solution of this system %for $V_k$
has the form
\begin{equation}
V_{r}=\sum_{p=1}^{n}\alpha_{rp}a_{p},\qquad \alpha_{rp}=\frac{\det(S^{(rp)})%
}{\det S}, \label{0.10}%
\end{equation}
where $S^{(rp)}$ is the matrix $S$
with the $r$-th column replaced by $(\lambda
_{1}S^p_1(\lambda_1\mu_1),\dots,$ $\lambda
_{n}S^p_n(\lambda_n\mu_n))^T$, the string of coefficients at the parameter $a_{p}$.
On the other hand, as $V=(V_{1},\dots,V_{n})^{T}$ and
$(a_{1},\dots,a_{n})^{T}$, system (\ref{0.9}) can be written in the matrix form as
\[
JV=\Lambda_{n}Ja\ \Longrightarrow\ V=J^{-1}\Lambda_{n}J\,a=\alpha\,a,
\]
where $\alpha_{ij}=(J^{-1}\Lambda_{n}J)_{ij}.$ Comparing this result with
(\ref{0.8}) and (\ref{0.10}) we find
\begin{equation}
F_{ij}=(J^{-1}\Lambda_{n}J)_{ij}=\frac{\det(S^{(ij)})%
}{\det S}. \label{0.11}%
\end{equation}

Now, the important question is which entries $F_{ij}$ are nonzero when the separation relations
take the form (\ref{6}). In other words, we want to know for which $i,j$
$\det(S^{(ij)})\neq 0$, i.e., the matrix $S^{(ij)}$ has no linearly dependent columns.

To answer this question, we first rewrite the quasi-bi-Hamiltonian
chain (\ref{0.7}) in the equivalent form
\begin{equation}
\Pi_{1}dH_{i}^{(k)}=\sum_{l=1}^{m}\sum_{j=1}^{n_{l}}F_{i,j}^{k,l}\,\Pi
_{0}\,dH_{j}^{(l)},\qquad k=1,\dots,m,\quad i=1,\dots,n_{k} \label{0.12}%
\end{equation}
adapted to the separation relations written in the form (\ref{9}). Then a simple
inspection shows that
\[
F_{i,i+1}^{k,k}=1,\qquad  F_{i,1}^{k,l}\equiv F_{i}^{k,l}\neq 0.
\]
Hence, the quasi-bi-Hamiltonian representation (\ref{0.12}) takes the form
\begin{equation}
\Pi_{1}dH_{i}^{(k)}=\Pi_{0}\,dH_{i+1}^{(k)}+\sum_{l=1}^{m}F_{i}^{k,l}%
\,\Pi_{0}\,dH_{1}^{(l)},\qquad H_{n_{k}+1}^{(k)}=0, \label{0.13}%
\end{equation}
where
\[
F_{i}^{k,l}=\frac{\det S_{i}^{(k,l)}}{\det S},
\]
and $S_{i}^{(k,l)}$ is the St\"{a}ckel matrix $S$ with
$(n_{1}+\dots+n_{k-1}+i)$-th column replaced by
$(\varphi_1^l\lambda_1^{n_l},\dots,\varphi_n^l\lambda_n^{n_l})^{T}$.

In the rest of this section we show that the representation
(\ref{0.13}) can be lifted to a GZ bi-Hamiltonian form. First, we
extend the $2n$ dimensional phase space
$M$ to $\mathcal{M}=M\times\mathbb{R}^m$
%\underset{m}{\underbrace{\mathbb{R}\times\dots\times\mathbb{R}}}$
with additional
coordinates $c_{i}$, $i=1,\dots,m$, on $\mathbb{R}^m$. Then, we extend the
Hamiltonians as follows:
\begin{equation}
H_{i}^{(k)}(q,p)\rightarrow h_{i}^{(k)}(q,p,c)=H_{i}^{(k)}(q,p)-\sum_{l=1}%
^{m}F_{i}^{k,l}(q,p)\,c_{l}. \label{0.14}%
\end{equation}
From (\ref{0.9})-(\ref{0.11})we infer that the separation relations for
$h_{i}^{(k)}$ read
\begin{equation}
\sum_{k=1}^{m}\varphi_i^k(\lambda_i,\mu_i)h^{(k)}(\lambda_i)=\psi_i
(\lambda_i,\mu_i),\qquad i=1,\dots,n,
\label{0.15}%
\end{equation}
where
\[
h^{(k)}(\lambda)=\sum_{i=0}^{n_{k}}\lambda^{n_{k}-i}h_{i}^{(k)}%
,\qquad h_{0}^{(k)}=c_{k},\qquad n_{1}+\dots+n_{m}=n.
\]

Moreover, for the functions $F_{i}^{k,l}$ we have the same quasi-bi-Hamiltonian
representation as for $H_{i}^{(k)}$:
\begin{equation}
\Pi_{1}dF_{i}^{k,l}=\Pi_{0}\,dF_{i+1}^{k,l}+\sum_{r=1}^{m}F_{i}%
^{k,r}\,\Pi_{0}\,dF_{1}^{r,l}, \label{0.16}%
\end{equation}
This was proved for arbitrary St\"{a}ckel systems in \cite{m3}.

Denote the push-forwards of the Poisson tensors $\Pi_{0}$ and $\Pi_{1}$ to $\mathcal{M}$
by $\pi_{0}$ and $\pi_{1D}$. %, respectively.
Both $\pi_{0}$ and $\pi_{1D}$ are
degenerate and possess common Casimirs $c_{i}$, $i=1,\dots,m$. We have
\begin{equation}
\pi_{0}=\left(
\begin{array}
[c]{c|c}%
\Pi_{0} & 0\\\hline
0 & 0
\end{array}
\right), \qquad \pi_{1D}=\left(
\begin{array}
[c]{c|c}%
\Pi_{1} & 0\\\hline
0 & 0
\end{array}
\right)  . \label{0.17}%
\end{equation}
Relations (\ref{0.13})-(\ref{0.17}) imply that on $\mathcal{M}$ we
have a quasi-bi-Hamiltonian representation with respect to the Poisson tensors
$\pi_{0}$ and $\pi_{1D}$ of the form
\begin{equation}
\pi_{1D}\,dh_{i}^{(k)}=\pi_{0}\,dh_{i+1}^{(k)}+\sum_{l=1}^{m}F_{i}%
^{k,l}\,\pi_{0}\,dh_{1}^{(l)},\qquad F_{0}^{k,l}=-\delta
_{kl},\quad h_{n_{k}+1}^{(k)}=0. \label{0.18}%
\end{equation}

Now introduce the bivector
\[
\pi_{1}:=\pi_{1D}+\sum_{k=1}^{m}X_{1}^{(k)}\wedge Z_{k},
\]
where
\[
X_{1}^{(k)}=\pi_{0}dh_{1}^{(k)},\quad Z_{k}=\tfrac{\partial}{\partial
c_{k}}.
\]

First we show that the bivector $\pi_{1}$ is Poisson.
Using the properties of the Schouten bracket we have
\begin{equation}
\lbrack\pi_{1},\pi_{1}]_{S}=2\sum_{i}Z_{i}\wedge L_{X_{1}^{(i)}}\pi_{1D}%
+2\sum_{i,j}[X_{1}^{(i)},Z_{j}]\wedge Z_{i}\wedge X_{1}^{(j)}, \label{0.18a}%
\end{equation}
where $L_{X}$ means the Lie derivative in the direction of $X$, and
$[\cdot,\cdot]$ is the commutator of vector fields. Now, let us prove that
\begin{equation}
L_{X_{1}^{(r)}}\Pi_{1D}=\sum_{l}\pi_{0}\,dF_{1}^{r,l}\wedge X_{1}^{(l)}.
\label{0.19}%
\end{equation}
From (\ref{0.18}) we have
\[
Y_{k}:=\pi_{1D}\,dh_{n_{k}}^{(k)}=\sum_{l}F_{n_{k}}^{k,l}X_{1}^{(l)},
\]%
\[
\pi_{1D}\,dF_{n_{k}}^{k,l}=\sum_{r}F_{n_{k}}^{k,r}\pi_{0}\,dF_{1}%
^{r,l}.
\]
From the Poisson property of $\pi_{1D}$ it follows that %we have
\begin{align*}
0  &  =L_{Y_{k}}\pi_{1D}\\
&  =\sum_{l}(F_{n_{k}}^{k,l}L_{X_{1}^{(l)}}\pi_{1D}-\pi_{1D}\,dF_{n_{k}%
}^{k,l}\wedge X_{1}^{(l)})\\
&  =\sum_{l}\left(  F_{n_{k}}^{k,l}L_{X_{1}^{(l)}}\pi_{1D}-(\sum_{r}%
F_{n_{k}}^{k,r}\pi_{0}\,dF_{1}^{r,l})\wedge X_{1}^{(l)}\right) \\
&  =\sum_{r}F_{n_{k}}^{k,r}L_{X_{1}^{(r)}}\pi_{1D}-\sum_{r}F_{n_{k}}%
^{k,r}\sum_{l}\pi_{0}\,dF_{1}^{r,l}\wedge X_{1}^{(l)}\\
&  =\sum_{r}F_{n_{k}}^{k,r}\left(  L_{X_{1}^{(r)}}\pi_{1D}-\sum_{l}\pi
_{0}\,dF_{1}^{r,l}\wedge X_{1}^{(l)}\right),
\end{align*}
hence (\ref{0.19}) is satisfied. On the other hand,
\[
\lbrack X_{1}^{(i)},Z_{j}]=\pi_{0}\,dF_{1}^{i,j},
\]
so (\ref{0.18a}) becomes
\[
\lbrack\pi_{1},\pi_{1}]_{S}=2\sum_{i,j}Z_{i}\wedge\pi_{0}\,dF_{1}%
^{i,j}\wedge X_{1}^{(j)}+2\sum_{i,j}\pi_{0}\,dF_{1}^{i,j}\wedge Z_{i}\wedge
X_{1}^{(j)}=0,
\]
and thus $\pi_1$ is Poisson.

Moreover, the Poisson bivectors $\pi_{0}$ and $\pi_{1}$ are compatible as
\[
\lbrack\pi_{0},\pi_{1}]_{S}=\sum_{i}(Z_{i}\wedge L_{X_{1}^{(i)}}\Pi_{0}%
-X_{1}^{(i)}\wedge L_{Z_{i}}\pi_{0})=0.
\]
Finally, the vector fields $X_{i}^{(k)}$ form bi-Hamiltonian chains with respect to
$\pi_{0},\pi_{1}$. Indeed, we have
\begin{align*}
\pi_{1}\,dh_{i}^{(k)}  &  =\pi_{0}\,dh_{i+1}^{(k)}+\sum_{l=1}^{m}F_{i}%
^{k,l}\,\pi_{0}\,dh_{1}^{(l)}+\sum_{l=1}^{m}(X_{1}^{(l)}\wedge Z_{l}%
)dh_{i}^{(k)}\\
&  =\pi_{0}\,dh_{i+1}^{(k)}=X_{i+1}^{(k)},
\end{align*}
as
\[
\sum_{l=1}^{m}(X_{1}^{(l)}\wedge Z_{l})dh_{i}^{(k)}=-\sum_{l=1}^{m}%
F_{i}^{k,l}\,X_{1}^{(l)}=-\sum_{l=1}^{m}F_{i}^{k,l}\Pi_{0}\,dh_{1}^{(l)}.
\]
Quite obviously,
\[
\pi_{1}\,dh_{n_{k}}^{(k)}=0,\qquad X_{1}^{(l)}=\pi_{1}\,dc_{l}=\Pi
_{1}\,dh_{0}^{(l)},
\]
so $h^{(k)}(\lambda)$ are polynomial in
%the pencil parameter
$\lambda$ Casimir functions of the Poisson pencil
$\pi_{\lambda}=\pi_{1}-\lambda\pi_{0}$.

\section{Examples}

Here we illustrate the above results with three examples of separable systems with three degrees of freedom.
Two of them are classical St\"{a}ckel systems with separation relations
quadratic in the momenta while the third example has separation relations cubic in the momenta.\\
\textbf{Example 1.}\\
Consider the separation relations
on a six-dimensional phase space
given by the following bare (potential-free) separation curve
\begin{equation*}
H_{1}\lambda ^{2}+H_{2}\lambda +H_{3}=\tfrac{1}{8}\mu ^{2}
\label{7.21a}
\end{equation*}
from the class (\ref{ben}). This curve corresponds to geodesic motion for
a classical St\"{a}ckel system (of Benenti type).
The transformation $(\lambda ,\mu)\rightarrow (q,p)$ to the flat coordinates of associated metric
follows from the point transformation
\begin{equation*}
\sigma_1(q)=q_{1}=\lambda ^{1}+\lambda ^{2}+\lambda ^{3},\ \ \sigma_2(q)=\tfrac{1}{4}q_{1}^{2}+%
\tfrac{1}{2}q_{2}=\lambda ^{1}\lambda ^{2}+\lambda ^{1}\lambda ^{3}+\lambda
^{2}\lambda ^{3},\ \  \sigma_3(q)=\tfrac{1}{4}q_{1}q_{2}+\tfrac{1}{4}q_{3}=\lambda
^{1}\lambda ^{2}\lambda ^{3}.
\end{equation*}
In the flat coordinates the Hamiltonians take the form
\begin{align*}
H_{1}& =p_{1}p_{3}+\tfrac{1}{2}p_{2}^{2},\\ \nonumber
H_{2}& =\tfrac{1}{2}q_{3}p_{3}^{2}-\tfrac{1}{2}q_{1}p_{2}^{2}+\tfrac{1}{2}%
q_{2}p_{2}p_{3}-\tfrac{1}{2}p_{1}p_{2}-\tfrac{1}{2}%
q_{1}p_{1}p_{3}, \\ \nonumber
H_{3}& =\tfrac{1}{8}q_{2}^{2}p_{3}^{2}+\tfrac{1}{8}q_{1}^{2}p_{2}^{2}+%
\tfrac{1}{8}p_{1}^{2}+\tfrac{1}{4}q_{1}p_{1}p_{2}+\tfrac{1}{4}%
q_{2}p_{1}p_{3}-\tfrac{1}{4}q_{1}q_{2}p_{2}p_{3}-\tfrac{1}{2}q_{3}p_{2}p_{3} \nonumber
\end{align*}%
and admit a quasi-bi-Hamiltonian representation (\ref{0.7}) with the operators $\Pi_{0}$ and $\Pi_{1}$
of the form
\begin{equation}
\Pi _{0}=\left(
\begin{array}{cc}
0 & I_{3} \\
-I_{3} & 0%
\end{array} \label{p0}
\right),
\end{equation}%
\begin{equation}
\Pi _{1}=\tfrac{1}{2}\left(
\begin{array}{cccccc}
0 & 0 & 0 & q_{1} & -1 & 0 \\
0 & 0 & 0 & q_{2} & 0 & -1 \\
0 & 0 & 0 & 2q_{3} & q_{2} & q_{1} \\
-q_{1} & -q_{2} & -2q_{3} & 0 & p_{2} & p_{3} \\
1 & 0 & -q_{2} & -p_{2} & 0 & 0 \\
0 & 1 & -q_{1} & -p_{3} & 0 & 0%
\end{array} \label{p1}
\right)
\end{equation}
and the control matrix
\[
F=\left(
\begin{array}{ccc}
q_1 & 1 & 0  \\
 -\tfrac{1}{4}q_1^2-\tfrac{1}{2}q_2^2& 0 & 1  \\
 \tfrac{1}{2}q_1q_2+\tfrac{1}{4}q_3& 0 & 0  \\
\end{array}\right).
\]
On the extended phase space of dimension seven, with an additional coordinate $c$,
we have the extended Hamiltonians (\ref{0.14})
\begin{align*}
h_0& =c,\\ \nonumber
h_{1}& =p_{1}p_{3}+\tfrac{1}{2}p_{2}^{2}-cq_1,\\ \nonumber
h_{2}& =\tfrac{1}{2}q_{3}p_{3}^{2}-\tfrac{1}{2}q^{1}p_{2}^{2}+\tfrac{1}{2}%
q_{2}p_{2}p_{3}-\tfrac{1}{2}p_{1}p_{2}-\tfrac{1}{2}%
q_{1}p_{1}p_{3}+(\tfrac{1}{4}q_1^2+\tfrac{1}{2}q_2^2)c, \\ \nonumber
h_{3}& =\tfrac{1}{8}q_{2}^{2}p_{3}^{2}+\tfrac{1}{8}q_{1}^{2}p_{2}^{2}+%
\tfrac{1}{8}p_{1}^{2}+\tfrac{1}{4}q_{1}p_{1}p_{2}+\tfrac{1}{4}%
q_{2}p_{1}p_{3}-\tfrac{1}{4}q_{1}q_{2}p_{2}p_{3}-\tfrac{1}{2}q_{3}p_{2}p_{3}-(\tfrac{1}{2}q_1q_2+\tfrac{1}{4}q_3)c. \label{7h}
\end{align*}
They form a bi-Hamiltonian chain
\begin{equation*}
\begin{array}{l}
\pi _{0}dh_{0}=0 \\
\pi _{0}dh_{1}=X_{1}=\pi _{1}dh_{0} \\
\pi _{0}dh_{2}=X_{2}=\pi _{1}dh_{1} \\
\pi _{0}dh_{3}=X_{3}=\pi _{1}dh_{2} \\
\qquad \quad \quad \,\,\, 0=\pi _{1}dh_{3,}\,\,%
\end{array}
\end{equation*}
with the Poisson operators $\pi _{0}$ and $\pi _{1}$, where
\begin{equation*}
\pi _{0}=\left(
\begin{array}{c|c}
\Pi _{0} & 0 \\ \hline
0 & 0%
\end{array}%
\right)\quad ,\quad \pi _{1}=\left(
\begin{array}{c|c}
\Pi_1 & X_{1} \\ \hline
-X_{3}^T & 0%
\end{array}%
\right).
\end{equation*}
The separation curve for the extended system takes the form
\[
c\lambda^3+h_{1}\lambda ^{2}+h_{2}\lambda +h_{3}=\tfrac{1}{8}\mu ^{2}.
\]
\textbf{Example 2.}\\
Consider now separation relations on a six-dimensional phase space given by the following
bare separation curve
\begin{equation*}
\bar{H_{1}}\lambda ^{3}+\bar{H_{2}}\lambda^2 +\bar{H_{3}}=\tfrac{1}{8}\mu ^{2}
\end{equation*}
from the class (\ref{99}).
When written using the notation (\ref{9}), this curve takes the form
\begin{equation*}
\lambda^2(H_{1}^{(1)}\lambda +H_{2}^{(1)}) +H_{1}^{(2)}=\tfrac{1}{8}\mu ^{2}
\label{7.21c}
\end{equation*}
and again represents geodesic motion for a classical St\"{a}ckel system (this time
of non-Benenti type).
Using the coordinates, the Hamiltonians, and the functions $\sigma_i$
from the previous example we find that
\begin{eqnarray*}
\bar{H}_{1}=H_{1}^{(1)} &=&-\tfrac{1}{\sigma _{2}}H_{2},  \notag \\
\bar{H}_{2}=H_{2}^{(1)} &=&H_{1}-\tfrac{\sigma _{1}}{\sigma _{2}}H_{2},   \\
\bar{H}_{3}=H_{1}^{(2)} &=&H_{3}-\tfrac{\sigma _{3}}{\sigma _{2}}H_{2} \notag  \\
\end{eqnarray*}
and thus we see that the Hamiltonians $\bar H_i$ are related to $H_i$ through the
so-called generalized St\"{a}ckel transform (see \cite{bs} for further details on the latter).
One can show that the metric tensor associated to the Hamiltonian $\bar{H_{1}}$ is not flat anymore.

The Hamiltonians $\bar H_i$ form a quasi-bi-Hamiltonian chain (\ref{0.7})
with the Poisson tensors (\ref{p0}),(\ref{p1})
and the control matrix
\[
F=\left(
\begin{array}{ccc}
\sigma_1-\tfrac{\sigma_3}{\sigma_2} & 1 & -\tfrac{1}{\sigma_2}  \\
 -\sigma_2+\tfrac{\sigma_1\sigma_3}{\sigma_2}& 0 & \tfrac{\sigma_1}{\sigma_2}  \\
 \tfrac{\sigma_3^2}{\sigma_2}& 0 & \tfrac{\sigma_3}{\sigma_2}  \\
\end{array}\right).
\]

On the extended phase space of dimension eight, with additional coordinates $c_1,c_2$,
we have the extended Hamiltonians (\ref{0.14})
\begin{eqnarray*}
h_0^{(1)} &=& c_1,\\
h_1^{(1)} &=& H_1^{(1)}-(\sigma_1-\tfrac{\sigma_3}{\sigma_2})c_1+\tfrac{1}{\sigma_2}c_2,\\
h_2^{(1)} &=& H_2^{(1)}+(\sigma_2-\tfrac{\sigma_1\sigma_3}{\sigma_2})c_1-\tfrac{\sigma_1}{\sigma_2}c_2,\\
h_0^{(2)} &=& c_2,\\
h_1^{(2)} &=& H_1^{(2)}-\tfrac{\sigma_3^2}{\sigma_2}c_1-\tfrac{\sigma_3}{\sigma_2}c_2.\\
\end{eqnarray*}
They form two bi-Hamiltonian chains
\begin{equation*}
\begin{array}{ccc}
\begin{array}{l}
\pi_{0}dh_{0}=0 \\
\pi_{0}dh_{0}^{(1)}=X_{1}^{(1)}=\pi_{1}dh_{0}^{(1)} \\
\pi_{0}dh_{2}^{(1)}=X_{2}^{(1)}=\pi_{1}dh_{1}^{(1)} \\
\qquad \qquad \qquad 0=\pi_{1}dh_{2}^{(1)}\,\,%
\end{array}
&  &
\begin{array}{l}
\pi_{0}dh_{0}^{(2)}=0 \\
\pi_{0}dh_{1}^{(2)}=X_{1}^{(2)}=\pi_{1}dh_{0}^{(2)} \\
\qquad \qquad \qquad 0=\pi_{1}dh_{1}^{(2)},%
\end{array}%
\end{array}
\end{equation*}
with the Poisson operators $\pi _{0}$ and $\pi _{1}$ of the form
\begin{equation*}
\pi_{0}=\left(
\begin{array}{c|c}
\Pi _{0} & 0\ \ 0 \\ \hline
\begin{array}{c}
0 \\
0
\end{array}
& 0%
\end{array}%
\right) \text{ \ , \ }\pi_{1}=\left(
\begin{array}{c|c}
\Pi_{1} & X_1^{(1)}\ \ \ X_1^{(2)} \\ \hline
\begin{array}{c}
-(X_1^{(1)})^{T} \\
-(X_1^{(2)})^{T}
\end{array}
& 0%
\end{array}%
\right) .
\end{equation*}
The separation curve for the extended system takes the form
\begin{equation*}
\lambda^2(c_1\lambda^2+h_{1}^{(1)}\lambda +h_{2}^{(1)}) +c_2\lambda+h_{1}^{(2)}=\tfrac{1}{8}\mu ^{2}.
\label{7.21c1}
\end{equation*}
\textbf{Example 3.}\\
Consider separation relations on a six-dimensional phase space
given by the following bare separation curve cubic in the momenta
\begin{equation*}
\mu H_1^{(1)} +H_{1}^{(2)}\lambda +H_{2}^{(2)}=\mu ^{3}
\end{equation*}
from the class (\ref{bus}). The transformation $(\lambda, \mu)\rightarrow (q,p)$ to new canonical coordinates
in which all Hamiltonians are of a polynomial form is obtained from the following two transformations:
\begin{eqnarray*}
u_1 &=& 3q_2-3q_3,\\
u_2 &=& -q_1p_2-q_1p_3+3q_3^2+5q_1^3-6q_2q_3,\\
u_3 &=& -q_3^3-9q_1^3q_3+q_1q_3p_2+q_1q_3p_3-\tfrac{2}{27}q_1^3q_2+q_1^2p_1+3q_2q_3^2,\\
v_1 &=& -\frac{1}{q_1},\\
v_2 &=& \frac{3q_2-2q_3}{q_1},\\
v_3 &=& p_3+\tfrac{2}{3}p_2-\frac{q_3^2}{q_1}+3\frac{q_2q_3}{q_1}-4q_1^2,
\end{eqnarray*}
and
\begin{eqnarray*}
u_1 &=& \lambda_1+\lambda_2+\lambda_3,\\
u_2 &=& \lambda_1\lambda_2+\lambda_1\lambda_3+\lambda_2\lambda_3,\\
u_3 &=& \lambda_1\lambda_2\lambda_3,\\
\mu_i &=& v_1\lambda_i^2+v_2\lambda_i+v_3, \qquad i=1,2,3.
\end{eqnarray*}

In the $(q,p)$-coordinates the Hamiltonians take the form
\begin{eqnarray*}
H_{1}^{(1)} & =& p_2p_3+\tfrac{1}{3}p_2^2+p_3^2-7q_1^2p_3-4q_1^2p_2-3q_2p_1+18q_1q_2^2+13q_1^4+12q_3q_1q_2,\\
H_{1}^{(2)} & =& 12q_1^3q_2+8q_1^3q_3-2q_1^2p_1+(-6q_1q_2-4q_1q_3)p_3+p_1p_3, \\
H_{2}^{(2)} & =& \tfrac{1}{3}p_2p_3^2+\tfrac{1}{3}p_2^2p_3+\tfrac{2}{27}p_2^3-q_1^2p_3^2-\tfrac{4}{3}q_1^2p_2^2-q_2p_1p_2-q_1p_1^2
 -\tfrac{10}{3}q_1^2p_3p_2\\
 &&+(q_3-3q_2)p_1p_3+(21q_1^2q_2+6q_3q_1^2)p_1
 +(4q_3q_1q_2+6q_1q_2^2
+\tfrac{22}{3}q_1^4)p_2\\
&&+(7q_1^4+18q_1q_2^2+6q_3q_1q_2-4q_1q_3^2)p_3-8q_1^3q_3^2-72q_3q_1^3q_2-90q_1^3q_2^2-12q_1^6. \label{77}
\end{eqnarray*}
They form a quasi-bi-Hamiltonian chain (\ref{0.7}) with the non-canonical Poisson operator
\begin{equation*}
\Pi_{1}=\left(
\begin{array}{cccccc}
0 & 0 & 0 & -q_{3} & 3q_1 & 2q_2 \\
0 & 0 & -\tfrac{1}{3}q_1 & A & 3q_2-q_3 & -q_2 \\
0 & \tfrac{1}{3}q_1 & 0 & 2q_{1}^2 & 0 & -q_{3} \\
-q_{3} & -A & -2q_{1}^2 & 0 & B & C \\
1-3q_1& -3q_2+q_3 & 0 & -B & 0 & -24q_1^2 \\
2q_1& q_2 & q_{3} & -C & 24q_1^2 & 0%
\end{array} \label{p11}
\right),
\end{equation*}
where $A=-\tfrac{1}{3}p_2+\tfrac{1}{3}p_3-3q_1^2$, $B=54q_1q_2+24q_1q_3-3p_1$, $C=-24q_1q_2-12q_1q_3+p_1$
and the control matrix
\[
F=\left(
\begin{array}{ccc}
-q_3 & -q_1 & 0  \\
 -\tfrac{1}{3}p_2+q_1^2& -2q_3+3q_2 & 1  \\
 5q_3q_1^2+6q_1^2q_2-q_1p_1-\tfrac{1}{3}q_3p_2& \, -4q_1^3-q_3^2+3q_2q_3+\tfrac{2}{3}q_1p_2+q_1p_3 & 0  \\
\end{array}\right).
\]

On the extended phase space of dimension eight, with additional coordinates $c_1,c_2$,
the extended Hamiltonians (\ref{0.14}) are
\begin{eqnarray*}
h_0^{(1)} &=& c_1,\\
h_1^{(1)} &=& H_1^{(1)}+q_3c_1+q_1c_2,\\
h_0^{(2)} &=& c_2,\\
h_1^{(2)} &=& H_1^{(2)}+(\tfrac{1}{3}p_2-q_1^2)c_1+( 2q_3-3q_2 )c_2,\\
h_2^{(2)} &=& H_2^{(2)}-(5q_3q_1^2+6q_1^2q_2-q_1p_1-\tfrac{1}{3}q_3p_2)
c_1-( -4q_1^3-q_3^2+3q_2q_3+\tfrac{2}{3}q_1p_2+q_1p_3)c_2.\\
\end{eqnarray*}
They form two bi-Hamiltonian chains
\begin{equation*}
\begin{array}{ccc}
\begin{array}{l}
\pi_{0}dh_{0}^{(1)}=0 \\
\pi_{0}dh_{1}^{(1)}=X_{1}^{(1)}=\pi_{1}dh_{0}^{(1)} \\
\qquad \qquad \qquad 0=\pi_{1}dh_{1}^{(1)}\,\,,%
\end{array}
& &
\begin{array}{l}
\pi_{0}dh_{0}^{(2)}=0 \\
\pi_{0}dh_{1}^{(2)}=X_{1}^{(2)}=\pi_{1}dh_{0}^{(2)} \\
\pi_{0}dh_{2}^{(2)}=X_{2}^{(2)}=\pi_{1}dh_{1}^{(2)} \\
\qquad \qquad \qquad 0=\pi_{1}dh_{2}^{(2)}\,\,%
\end{array}
\end{array}
  \label{7.15q}
\end{equation*}
with the corresponding Poisson operators $\pi _{0}$ and $\pi _{1}$
\begin{equation*}
\pi_{0}=\left(
\begin{array}{c|c}
\Pi _{0} & 0\ \ 0 \\ \hline
\begin{array}{c}
0 \\
0
\end{array}
& 0%
\end{array}%
\right) \text{ \ , \ }\pi_{1}=\left(
\begin{array}{c|c}
\Pi_{1} & X_1^{(1)}\ \ \ X_1^{(2)} \\ \hline
\begin{array}{c}
-(X_1^{(1)})^{T} \\
-(X_1^{(2)})^{T}
\end{array}
& 0%
\end{array}%
\right) .
\end{equation*}
The separation curve for the extended system takes the form
\begin{equation*}
\mu(c_1\lambda+h_{1}^{(1)}) +c_2\lambda^2+h_{1}^{(2)}\lambda+h_{2}^{(2)}=\mu ^{3}.
\label{7.21e}
\end{equation*}

\section{Summary}

We have considered the St\"{a}ckel systems classified using their
separation relations. The most general form of the
separation relations considered in the present paper
is
\[
\sum_{k=1}^{m}S_i^{k}(\mu_{i},\lambda_{i})H^{(k)}(\lambda_{i})=\psi_{i}%
(\lambda_{i},\mu_{i}),\qquad m\leq n,\quad i=1,\dots,n,
\]
where
\[
H^{(k)}(\lambda_{i})=\sum_{j=1}^{n_{k}}\lambda_{i}^{n_{k}-j}H_{j}%
^{(k)},\qquad n_{1}+\dots+n_{m}=n
\]
and $S_i^{k},\psi_{i}$ are smooth functions of their arguments. Moreover, we have proved that all systems
whose separation relations are of the above form
admit (after the lift to an extended phase space) Gel'fand-Zakharevich bi-Hamiltonian
representation. This confirms universality of the latter property for the St\"{a}ckel
systems. As a consequence, a geometric separability theory,
based on the existence of GZ bi-Hamiltonian representation of a given system, is
applicable for all Liouville integrable systems from the classes we considered.

Quite obviously, the knowledge of quasi-bi-Hamiltonian representation is sufficient for separability of a given
Liouville integrable system. Unfortunately, there is no systematic method available for the construction
of such representation. On the other hand, there are some systematic
methods for finding the bi-Hamiltonian representation. From this point of view
the result presented in this paper is of interest, as it shows that the existence of bi-Hamiltonian
representation is an inherent property of St\"{a}ckel systems.

\section*{Acknowledgments}

This research was supported in part by the Ministry of
Science and Higher Education (MNiSW) of the Republic of Poland under
the research grant No.~N~N202~4049~33 and by the Ministry of Education, Youth and Sports of the Czech Republic
(M\v{S}MT \v{C}R) under grant MSM 4781305904.

The author appreciates warm hospitality of the Mathematical Institute of
Silesian University in Opava and stimulating discussions with Dr. A. Sergyeyev
on the subject of the present paper.

\end{document}